\documentclass[a4paper,12pt]{article}
\usepackage{graphicx}
\usepackage{epsfig}
\begin{document}
\title{Analytic solution for kinetic equilibrium of beta-processes in nucleonic plasma with relativistic pairs}

\author{G.S. Bisnovatyi-Kogan,
\thanks{ Space Research Institute,
Russian Academy of Sciences, Moscow, Russia.}}
\date{}
\maketitle

\begin{abstract}
The analytic solution is obtained describing kinetic equilibrium of the $\beta$-processes in the nucleonic plasma with relativistic pairs. The nucleons $(n,p)$ are supposed to be non-relativistic
and non-degenerate, while the electrons and positrons are ultra-relativistic due to high temperature $(T>6\cdot 10^9$K), or high density $(\rho>\mu 10^6$g/cm$^3$),
 or both, where $\mu$ is a number of nucleons per one electron. The consideration is simplified because of the analytic connection of the density with the electron
 chemical potential in the ultra-relativistic plasma, and Gauss representation of Fermi functions. Electron chemical potential and number of nucleons per one initial electron are calculated as functions of $\rho$ and $T$.
\end{abstract}

\maketitle

\section{\label{sec:level1} Introduction}

During a collapse, leading to a formation of a neutron star, the matter passes states with very high temperatures and densities, and intensive birth of neutrino. While in the hot core of the new born neutron star the neutrino opacity is large enough, to establish a thermodynamic equilibrium to beta processes, the regions outside the neutrinosphere are almost transparent to neutrino, so the thermodynamic equilibrium cannot be established. It was shown in \cite{inp67}, that the characteristic time of the neutrino processes outside the neutrinosphere can be much smaller than the characteristic hydrodynamic time. In this conditions the kinetic equilibrium is established, so that the ratio between neutrons and protons is determined by the conditions of the kinetic equilibrium, where the birth rate of protons (neutrons) in equal to its death rate. The detailed investigation of the kinetic equilibrium for the pure nucleonic $(n-p)$ gas had been done numerically in \cite{inp67} for a general case. Particular cases of the kinetic $\beta$ equilibrium in a cold $pne^-$, and a hot $pne^\pm$ gases have been considered numerically in \cite{prd05}, where the author described the kinetic $\beta$ equilibrium approximately, in terms of the relations between the chemical potentials of nucleon and pairs, similar to \cite{in65}.
The precision of this approach is rather moderate at high densities.

Here the same problem is considered for the case of ultrarelativistic pairs with non-relativistic and non-degenerate nucleons. This case is considered without any other simplifications, and occurs  to be rather simple, due to analytic connection of the matter density $\rho$ with the chemical potential of the electrons $\mu_e$, found in \cite{rho50}, see also \cite{nad74}.

\section{$\beta$ reaction rates}

Let us consider kinetic equilibrium for  ultrarelativistic pairs  to the
following processes

\begin{equation}
e^-+p \rightarrow n+\nu_e,\quad (a)
\label{rc}
\end{equation}
\[e^+ +n \rightarrow p+\tilde\nu_e. \quad(b)\]
The probabilities of the reactions (\ref{rc}) for infinitely heavy nucleons are written as \cite{inp67,bk01}

\begin{equation}
W^{\rm (a)}={\ln 2\over (Ft_{1/2}
)_{\rm n}} \left ( {kT\over m_{\rm e}c^2}\right )^5 I_2,
\label{pr}
\end{equation}
\[W^{\rm (b)}={\ln 2\over (Ft_{1/2}
)_{\rm n}} \left ( {kT\over m_{\rm e}c^2}\right )^5 J_2.\]
Here $(Ft_{1/2})_{\rm n}\approx 1200$ is the characteristic value for the neutron decay. The integrals $I_2$ and $J_2$ are defined as

\begin{equation}
I_2=\int\limits^\infty_0{x^2(x+x_0)\sqrt{(x+x_0)^2-\alpha^2}\ dx\over
1+\exp(x+x_0-\beta)},\qquad ({\rm a})
\label{int}
\end{equation}
\[J_2=\int\limits^\infty_0{(x+x_0+\alpha)^2(x+\alpha)
 \sqrt{x^2+2\alpha x}\ dx\over
1+\exp(x+\alpha+\beta)}. \qquad ({\rm b})\]
Here

\[x_0=\Delta_{np}/kT,\,\, \Delta_{np}=m_n-m_p\approx 1.29 \,{\rm MeV},\,\,\]
\begin{equation}
\alpha=m_e c^2/kT,\,\, \beta=\mu_e/kT.
\label{x0}
\end{equation}
For the positrons the chemical potential $\mu_{e^+}=-\mu_e$. In the upper integral (\ref{int}) let us define ($x=E_e/kT-x_0$), and in the lower one ($x=E_{e^+}/kT-\alpha$), ($E_{ee^+}=\sqrt{p^2 c^2+m_e c^2}$), $p$ is the electron(positron) momentum.

In the ultrarelativistic plasma $x+x_0 \gg \alpha$ in the case (a), and $x\gg \alpha$ in the case (b). Therefore the integrals (\ref{int}) are reduced to

\begin{equation}
I_2=\int\limits^\infty_0{x^2(x+x_0)^2\ dx\over
1+\exp(x+x_0-\beta)},\qquad ({\rm a})
\label{int1}
\end{equation}
\[J_2=\int\limits^\infty_0{(x+x_0)^2x^2
\ dx\over
1+\exp(x+\beta)}. \qquad ({\rm b})\]
Introducing Fermi integrals

\begin{equation}
F_n(\alpha)=\int_0^\infty\frac{x^n dx}{1+\exp(x-\alpha)},
\label{fi}
\end{equation}
let us write the integrals in (\ref{int1}) as

\begin{equation}
I_2=F_4(\beta-x_0)+2 x_0 F_3(\beta-x_0)+x_0^2 F_2(\beta-x_0), \qquad
\label{int2}
\end{equation}
\[J_2=F_4(-\beta)+2 x_0 F_3(-\beta)+x_0^2 F_2(-\beta).\]
In presence of ultrarelativistic pairs in thermodynamic equilibrium, there is an analytic expression, connecting the non-dimensional chemical potential $\beta$ with the matter density $\rho$, temperature $T$, and the number of nucleons per one original electron $\mu=\frac{n_p+n_n}{n_p}$ in the nucleonic medium, written as \cite{rho50,nad74,bk01}

\begin{equation}
{\rho\over \mu m_p}={1\over 3\pi^2}\left(kT\over \hbar c \right)^3
  (\beta^3+\pi^2\beta).
\label{eos}
\end{equation}

\section{Kinetic beta equilibrium}

\begin{figure}[htp]
\centerline{\hbox{\includegraphics[width=0.8\textwidth]{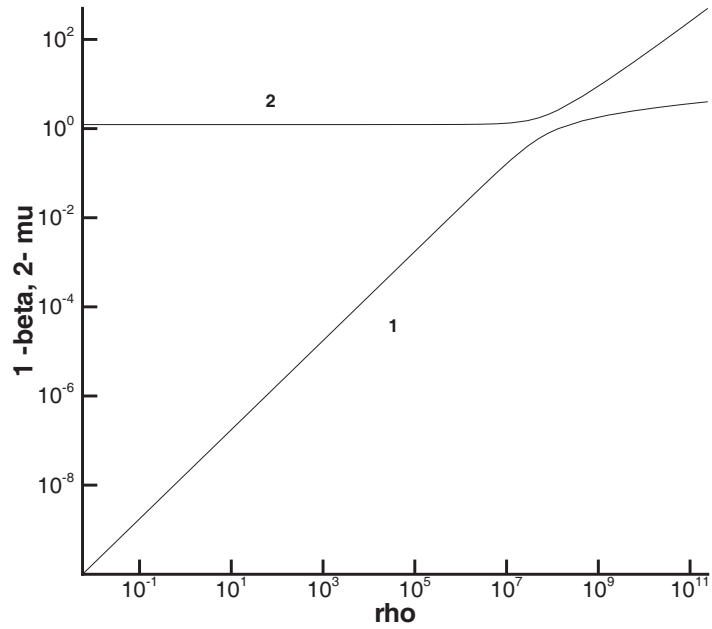}}}
\caption{The dependence  of the non-dimensional chemical potential
$\beta=\mu_e/kT$, and the number of nucleons per one initial
electron $\mu=1+\frac{n_n}{n_p}$, as a function density $\rho$ (g/cm$^3$), for
fixed values of temperature, $T=10^{10}$ K, in the kinetic beta
equilibrium with the ultrarelativistic pairs,  in the nucleonic
plasma, with $\Delta_{np}=1.293$ MeV (corresponding to $1.5\cdot
10^{10}$ K).} \label{fig1}
\end{figure}

\begin{figure}[htp]
\centerline{\hbox{\includegraphics[width=0.8\textwidth]{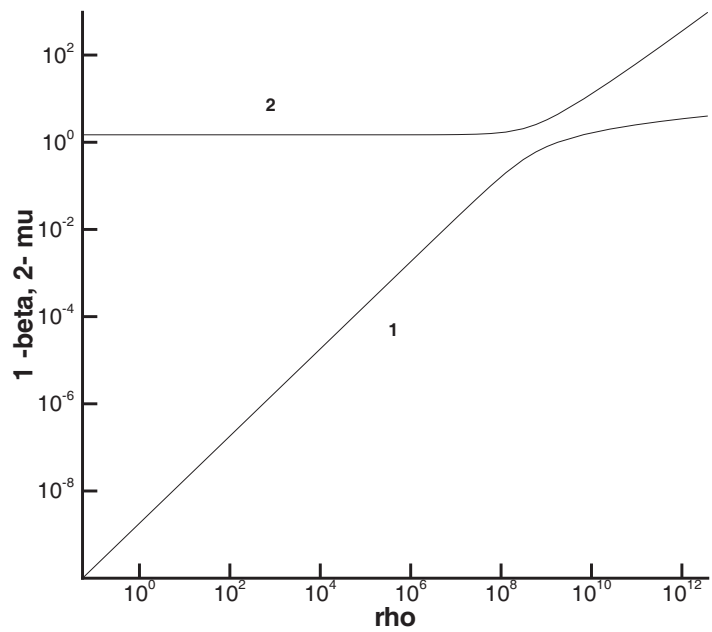}}}
\caption{Same as in Fig.\ref{fig1}, for for fixed values of temperature, for
fixed values of temperature, $T=2\cdot 10^{10}$ K.} \label{fig1a}
\end{figure}

\begin{figure}[htp]
\centerline{\hbox{\includegraphics[width=0.8\textwidth]{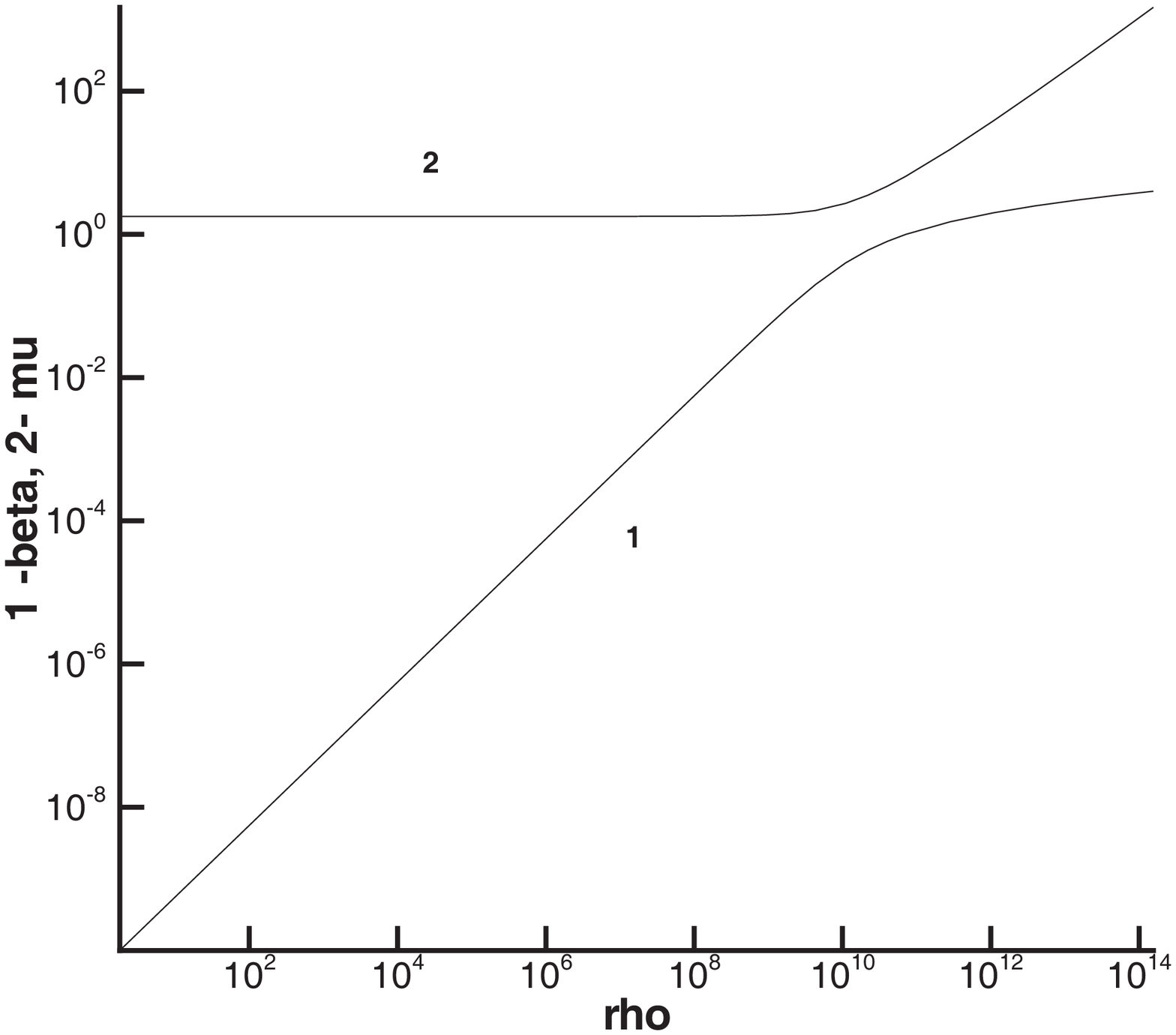}}} \caption{Same as in Fig.\ref{fig1}, for for fixed values of
temperature, $T=6\cdot 10^{10}$ K. For non-degenerate nucleons
the density should not exceed $1.3\cdot 10^{13}$ g/cm$^3$, according to
(\ref{nt}).} \label{fig2}
\end{figure}

\begin{figure}[htp]
\centerline{\hbox{\includegraphics[width=0.8\textwidth]{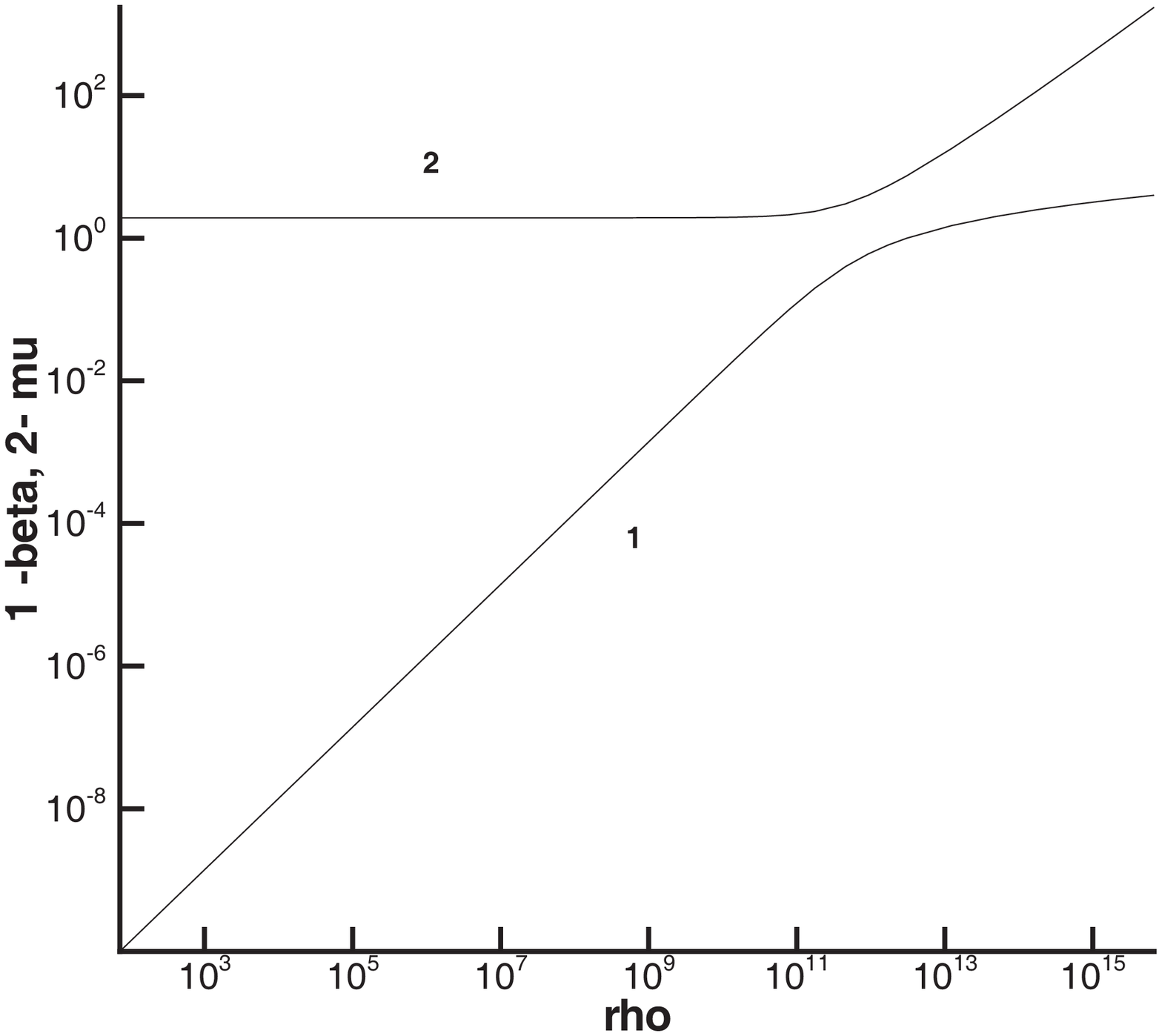}}} \caption{Same as in Fig.\ref{fig1}, for for fixed values of
temperature, $T=2\cdot 10^{11}$ K. For non-degenerate nucleons
the density should not exceed $8.1\cdot 10^{13}$ g/cm$^3$, according to
(\ref{nt}). } \label{fig2a}
\end{figure}

\begin{figure}[htp]
\centerline{\hbox{\includegraphics[width=0.8\textwidth]{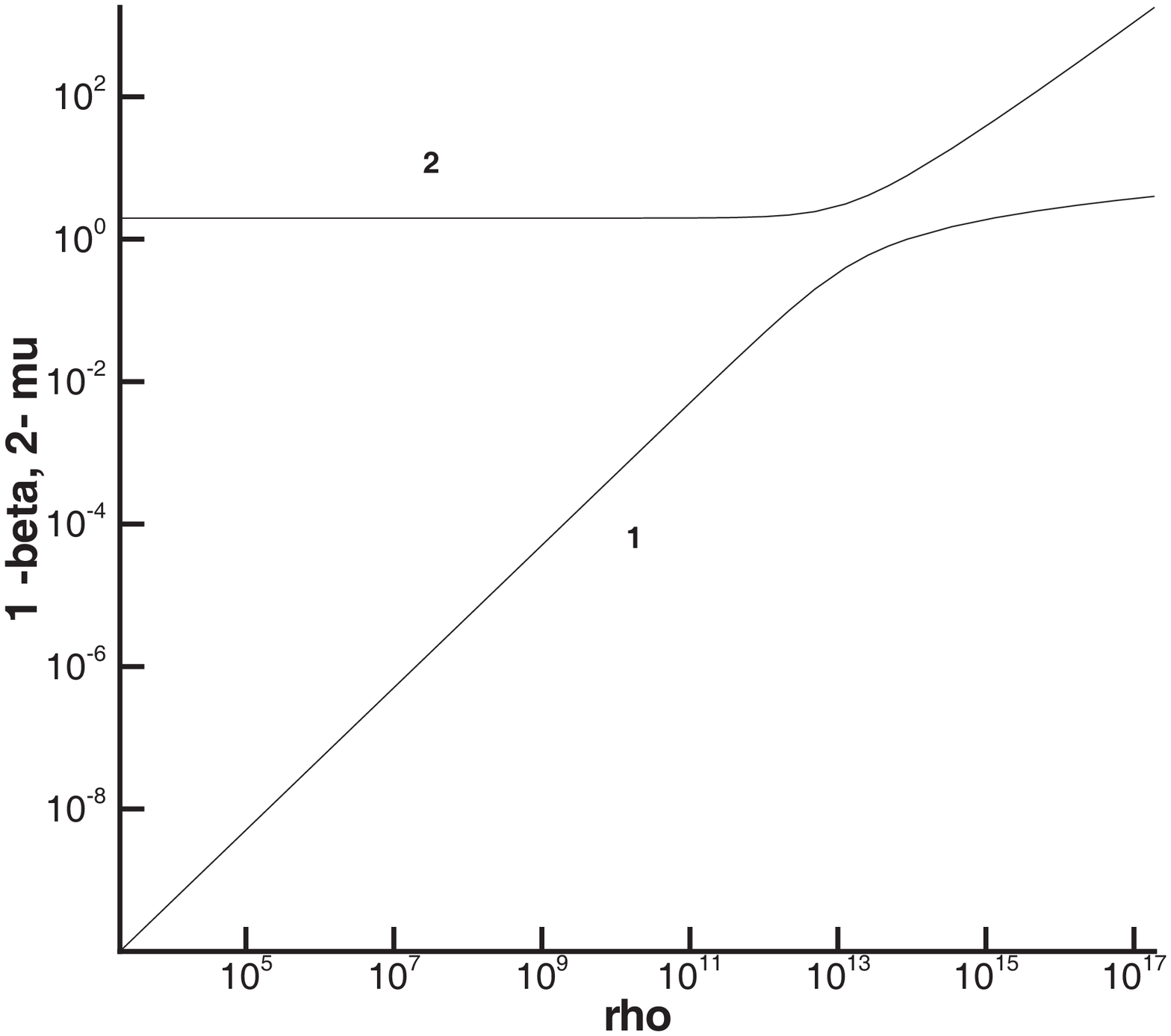}}}
\caption{Same as in Fig.\ref{fig1}, for for fixed values of
temperature, $T=6\cdot 10^{11}$ K. For non-degenerate nucleons
the density should not exceed $4.2\cdot 10^{14}$ g/cm$^3$, according to
(\ref{nt}).} \label{fig3}
\end{figure}

In presence of ultrarelativistic pairs the kinetic beta equilibrium is provided by the electron and positron captures, and the input from the neutron decay is negligible. In this conditions the equation of the kinetic beta equilibrium is written as

 \begin{equation}
n_p W^{(a)} = n_n W^{(b)},\quad \mu=1+\frac{W^{(a)}}{ W^{(b)}}=1+\frac{I_2}{J_2}.
\label{kbe}
\end{equation}
In the ultrarelativistic case thermodynamic functions as functions of $\rho $ and $T$ depend on the combination $\rho/T^3$. Introducing a non-dimensional variable

\begin{equation}
Z=3\pi^2\frac{\rho}{m_p}\left(\hbar c \over kT \right)^3,
\label{var}
\end{equation}
it follows from (\ref{eos}) and (\ref{kbe}) the equation, determining a dependence $\beta(Z,x_0)$ in the conditions of the kinetic beta equilibrium

 \begin{equation}
Z= (\beta^3+\pi^2\beta)\left(1+\frac{I_2}{J_2}\right),
\label{zet}
\end{equation}
where the integrals $I_2(\beta,x_0)$ and $J_2(\beta,x_0)$ are defined in (\ref{int2}).
The chemical composition, represented by the value of $\mu$ is determined than as

 \begin{equation}
\mu= \frac{Z}{\beta^3+\pi^2\beta}.
\label{mu}
\end{equation}
The dependences $\beta(\rho)|_{T}$ and $\mu(\rho)|_{T}$ in the kinetic beta equilibrium are presented in Figs. \ref{fig1} - \ref{fig3}.

The same dependences are represented in Tabs.\ref{tab1}-\ref{tab3}. The results represented in Figs. \ref{fig1} - \ref{fig3} and  Tabs.\ref{tab1}-\ref{tab3} coincide with numerical results from \cite{inp67}, presented in Figs 1-3 of this paper.

\begin{table}
\begin{center}
\caption{The dependence  of the non-dimensional chemical potential
$\beta=\mu_e/kT$, and the number of nucleons per one initial
electron $\mu=1+\frac{n_n}{n_p}$, as a function density $\rho$, for
fixed values of temperature, $T=10^{10}$ K (left), and $T=2\cdot
10^{10}$ K (right), in the kinetic beta equilibrium with the
ultrarelativistic pairs,  in the nucleonic plasma, with
$\Delta_{np}=1.293$ MeV (corresponding to $1.5\cdot 10^{10}$ K).}
\vspace{3mm}
 \label{tab1}
  \footnotesize{
\begin{tabular}{cllllllll}
\hline
T (K) & $\rho$(g/cm$^3$) & $\beta$ &$\mu$ &  T (K)& $\rho$(g/cm$^3$) & $\beta$ &$\mu$\\
\hline
 &&&&&&&& \\
$10^{10}$  & $5.72\cdot 10^3$ & $10^{-4}$   & 1.23 &$2\cdot 10^{10}$ & $5.50\cdot 10^4$ &$10^{-4}$ &1.48     \\
$10^{10}$  & $5.72\cdot 10^4$ & $10^{-3}$  & 1.23 &$2\cdot 10^{10}$ & $5.51 \cdot 10^5$ &$10^{-3}$ &1.48   \\
$10^{10}$  & $2.86\cdot 10^5$ & $0.005$   & 1.23 &$2\cdot 10^{10}$ & $2.76 \cdot 10^6$ &0.005 &1.49    \\
$10^{10}$  & $5.74\cdot 10^5$ & 0.01 & 1.24  &$2\cdot 10^{10}$ & $5.54 \cdot 10^6$ &0.01 &1.49 \\
$10^{10}$ & $1.15\cdot 10^6$  &$0.02$ & 1.24  &$2\cdot 10^{10}$ & $1.11\cdot 10^{7}$ &0.02 &1.50\\
$10^{10}$  & $2.91\cdot 10^6$  & $0.05$   &  1.25  &$2\cdot 10^{10}$ & $2.84\cdot 10^{7}$ &0.05 & 1.53   \\
$10^{10}$  & $5.95\cdot 10^6$  & 0.1  & 1.28   &$2\cdot 10^{10}$ &$5.89\cdot 10^{7}$ &0.1 &1.58    \\
$10^{10}$  & $1.25\cdot 10^7$  & 0.2   & 1.34 &$2\cdot 10^{10}$ &$1.28\cdot 10^{8}$ &0.2 &1.71   \\
$10^{10}$  & $2.84\cdot 10^7$   & 0.4 & 1.50 &$2\cdot 10^{10}$ &$3.10\cdot 10^{8}$ &0.4 &2.05     \\
$10^{10}$  & $5.04\cdot 10^7$   & 0.6  & 1.75 &$2\cdot 10^{10}$ & $5.89\cdot 10^{8}$ &0.6 &2.55     \\
$10^{10}$  & $8.32\cdot 10^7$   & 0.8  & 2.10 &$2\cdot 10^{10}$ & $1.04\cdot 10^{9}$ &0.8 &3.29     \\
$10^{10}$  & $1.35\cdot 10^8$   & 1   & 2.63 &$2\cdot 10^{10}$ & $1.79\cdot 10^{9}$ &1 &4.38    \\
$10^{10}$  & $4.56\cdot 10^8$   & 1.5  & 5.33 &$2\cdot 10^{10}$ & $6.78\cdot 10^{9}$ &1.5 &9.91    \\
$10^{10}$  & $1.62\cdot 10^9$   & 2  & 12.4 &$2\cdot 10^{10}$ & $2.54\cdot 10^{10}$ &2 &24.3     \\
$10^{10}$  & $5.87\cdot 10^9$   & 2.5  & 30.9 &$2\cdot 10^{10}$ & $9.30\cdot 10^{10}$ &2.5 &61.3     \\
$10^{10}$  & $2.09\cdot 10^{10}$   & 3   & 78.5 &$2\cdot 10^{10}$ & $3.31\cdot 10^{11}$ &3 &155     \\
$10^{10}$  & $7.25\cdot 10^{10}$   & 3.5   & 199 &$2\cdot 10^{10}$ & $1.13\cdot 10^{12}$ &3.5 &389    \\
$10^{10}$  & $2.43\cdot 10^{11}$   & 4   & 499 & &  & &    \\
 &&&&&&&& \\
\hline
\end{tabular}}
\end{center}
\end{table}

\begin{table}
\begin{center}
\caption{The same, as in Tab.\ref{tab1}, for  fixed values of the
temperature, $T=6\cdot 10^{10}$ K (left), and $T=2\cdot 10^{11}$ K
(right).} \vspace{3mm}
 \label{tab2}
\footnotesize{
\begin{tabular}{cllllllll}
\hline
T (K) & $\rho$(g/cm$^3$) & $\beta$ &$\mu$ &  T (K)& $\rho$(g/cm$^3$) & $\beta$ &$\mu$\\
\hline
 &&&&&&&& \\
$6\cdot 10^{10}$  & $1.79\cdot 10^6$ & $10^{-4}$   & 1.78 &$2\cdot 10^{11}$ & $7.17\cdot 10^7$ &$10^{-4}$ &1.93     \\
$6\cdot 10^{10}$  & $1.79\cdot 10^7$ & $10^{-3}$  & 1.79 &$2\cdot 10^{11}$ & $7.17 \cdot 10^8$ &$10^{-3}$ &1.93  \\
$6\cdot 10^{10}$  & $8.98\cdot 10^7$ & $0.005$   & 1.79 &$2\cdot 10^{11}$ & $3.60 \cdot 10^9$ &0.005 &1.94   \\
$6\cdot 10^{10}$  & $1.80\cdot 10^8$ & 0.01 & 1.80  &$2\cdot 10^{11}$ & $7.24 \cdot 10^9$ &0.01 &1.95 \\
$6\cdot 10^{10}$ & $3.64\cdot 10^8$  &$0.02$ & 1.82  &$2\cdot 10^{11}$ & $1.46\cdot 10^{10}$ &0.02 &1.97\\
$6\cdot 10^{10}$  & $9.35\cdot 10^8$  & $0.05$   &  1.86  &$2\cdot 10^{11}$ & $3.76\cdot 10^{10}$ &0.05 & 2.02  \\
$6\cdot 10^{10}$  & $1.96\cdot 10^9$  & 0.1  & 1.95   &$2\cdot 10^{11}$ &$7.92\cdot 10^{10}$ &0.1 &2.13   \\
$6\cdot 10^{10}$  & $4.34\cdot 10^9$  & 0.2   & 2.16 &$2\cdot 10^{11}$ &$1.77\cdot 10^{11}$ &0.2 &2.37  \\
$6\cdot 10^{10}$  & $1.10\cdot 10^{10}$   & 0.4 & 2.71 &$2\cdot 10^{11}$ &$4.57\cdot 10^{11}$ &0.4 &3.03     \\
$6\cdot 10^{10}$  & $2.20\cdot 10^{10}$   & 0.6  & 3.52 &$2\cdot 10^{11}$ & $9.21\cdot 10^{11}$ &0.6 &3.99    \\
$6\cdot 10^{10}$  & $4.03\cdot 10^{10}$   & 0.8  & 4.72 &$2\cdot 10^{11}$ & $1.71\cdot 10^{12}$ &0.8 &5.41    \\
$6\cdot 10^{10}$  & $7.16\cdot 10^{10}$   & 1   & 6.48 &$2\cdot 10^{11}$ & $3.06\cdot 10^{12}$ &1 &7.49    \\
$6\cdot 10^{10}$  & $2.84\cdot 10^{11}$   & 1.5  & 15.4 &$2\cdot 10^{11}$ & $1.23\cdot 10^{13}$ &1.5 &18.0   \\
$6\cdot 10^{10}$  & $1.08\cdot 10^{12}$   & 2  & 38.4 &$2\cdot 10^{11}$ & $4.71\cdot 10^{13}$ &2 &45.1     \\
$6\cdot 10^{10}$  & $3.98\cdot 10^{12}$   & 2.5  & 97.1 & & & &   \\
$6\cdot 10^{10}$  & $1.41\cdot 10^{13}$   & 3   & 245 & &  & &   \\
 &&&&&&&& \\
\hline
\end{tabular}}
\end{center}
\end{table}

\begin{table}
\begin{center}
\caption{The same, as in Tab.\ref{tab1}, for  fixed values of the
temperature, $T=6\cdot 10^{11}$ K (left), and $T=1.5\cdot 10^{12}$ K
(right).} \vspace{3mm}
 \label{tab3}
 \footnotesize{
 \begin{tabular}{cllllllll}
\hline
T (K) & $\rho$(g/cm$^3$) & $\beta$ &$\mu$ &  T (K)& $\rho$(g/cm$^3$) & $\beta$ &$\mu$\\
\hline
 &&&&&&&& \\
$6\cdot 10^{11}$  & $1.79\cdot 10^6$ & $10^{-5}$   & 1.78 &$1.5\cdot 10^{12}$ & $7.17\cdot 10^7$ &$10^{-5}$ &1.93     \\
$6\cdot 10^{11}$  & $1.79\cdot 10^7$ & $10^{-4}$  & 1.79 &$1.5\cdot 10^{12}$ & $7.17 \cdot 10^8$ &$10^{-4}$ &1.93  \\
$6\cdot 10^{11}$  & $8.98\cdot 10^7$ & $10^{-3}$   & 1.79 &$1.5\cdot 10^{12}$ & $3.60 \cdot 10^9$ &$10^{-3}$ &1.94   \\
$6\cdot 10^{11}$  & $1.80\cdot 10^8$ & 0.005 & 1.80  &$1.5\cdot 10^{12}$ & $7.24 \cdot 10^9$ &0.005 &1.95 \\
$6\cdot 10^{11}$ & $3.64\cdot 10^8$  &$0.01$ & 1.82  &$1.5\cdot 10^{12}$ & $1.46\cdot 10^{10}$ &0.01 &1.97\\
$6\cdot 10^{11}$  & $9.35\cdot 10^8$  & $0.02$   &  1.86  &$1.5\cdot 10^{12}$ & $3.76\cdot 10^{10}$ &0.02 & 2.02  \\
$6\cdot 10^{11}$  & $1.96\cdot 10^9$  & 0.05  & 1.95   &$1.5\cdot 10^{12}$ &$7.92\cdot 10^{10}$ &0.05 &2.13   \\
$6\cdot 10^{11}$  & $4.34\cdot 10^9$  & 0.1   & 2.16 &$1.5\cdot 10^{12}$ &$1.77\cdot 10^{11}$ &0.1 &2.37  \\
$6\cdot 10^{11}$  & $1.10\cdot 10^{10}$   & 0.2 & 2.71 &$1.5\cdot 10^{12}$ &$4.57\cdot 10^{11}$ &0.2 &3.03     \\
$6\cdot 10^{11}$  & $2.20\cdot 10^{10}$   & 0.4  & 3.52 &$1.5\cdot 10^{12}$ & $9.21\cdot 10^{11}$ &0.4 &3.99    \\
$6\cdot 10^{11}$  & $4.03\cdot 10^{10}$   & 0.6  & 4.72 &$1.5\cdot 10^{12}$ & $1.71\cdot 10^{12}$ &0.6 &5.41    \\
$6\cdot 10^{11}$  & $7.16\cdot 10^{10}$   & 0.8   & 6.48 &$1.5\cdot 10^{12}$ & $3.06\cdot 10^{12}$ &0.8 &7.49    \\
$6\cdot 10^{11}$  & $2.84\cdot 10^{11}$   & 1  & 15.4 &$1.5\cdot 10^{12}$ & $1.23\cdot 10^{13}$ &1 &18.0   \\
$6\cdot 10^{11}$  & $1.08\cdot 10^{12}$   & 1.5  & 38.4 &$1.5\cdot 10^{12}$ & $4.71\cdot 10^{13}$ &1.5 &45.1     \\
$6\cdot 10^{11}$  & $3.98\cdot 10^{12}$   & 2  & 97.1 &$1.5\cdot 10^{12}$ & $1.73\cdot 10^{14}$ &2 &114    \\
$6\cdot 10^{11}$  & $1.41\cdot 10^{13}$   & 2.5   & 245 &$1.5\cdot 10^{12}$ & $6.11\cdot 10^{14}$ &2.5 &287     \\
 &&&&&&&& \\
\hline
\end{tabular}}
\end{center}
\end{table}

\begin{figure}[htp]
\centerline{\hbox{\includegraphics[width=0.8\textwidth]{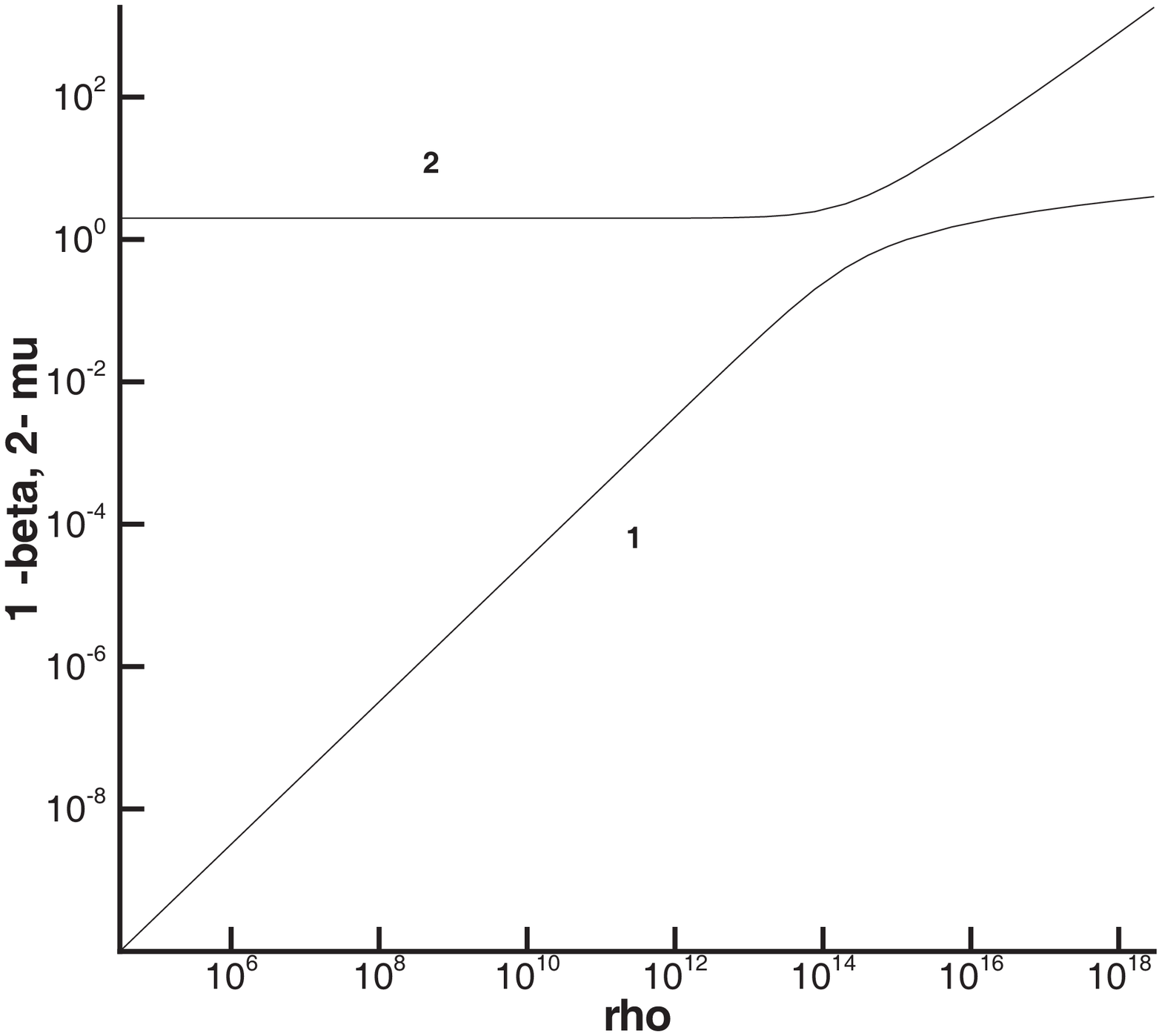}}} \caption{Same as in Fig.\ref{fig1}, for for fixed values of
temperature, $T=1.5\cdot 10^{12}$ K. For non-degenerate nucleons
the density should not exceed $1.7\cdot 10^{15}$ g/cm$^3$, according to
(\ref{nt}).} \label{fig3a}
\end{figure}

\section{Discussion}

Note, that the paper \cite{inp67} all integrals and algebraic
equations, for  the composition of the nucleonic plasma in kinetic
beta equilibrium, have been solved numerically for a general case,
while here a semi-analytical solution for the  ultrarelativistic
case is obtained. The dependence $\beta(\rho)$ was obtained in
\cite{prd05} using approximate approach, also for the
ultrarelativistic pair conditions, the consideration here is exact
for this case. The consideration for  the kinetic beta equilibrium
with ultrarelativistic pairs may be generalized for the mixture of
nuclei in a nuclear equilibrium. In this we have several parameters
$x_{zz'}=\Delta_{ZZ'}/kT$. The account of the analytic connection
between $\beta,\,\,\rho$ and $\mu$ in the ultrarelativistic
conditions, should simplify the calculations, performed in
\cite{chech69} numerically for the general case.

The kinetic beta equilibrium is applied to the regions around the
neutrinosphere, in core- collapse supernovae calculations. The
temperature $T_{\nu}$ and density $\rho_{\nu}$ at the neutrinosphere
have been calculated in \cite{n78}, giving

 \begin{equation}
T_{\nu}\,=\,4.5\cdot 10^{10}\,-\,6.3\cdot 10^{10}\,\,\,{\rm K},
 \label{nsph}
\end{equation}
\[
\rho_{\nu}\,=\,3\cdot 10^{11}\,-\,3\cdot 10^{12}\,\,\,{\rm
g/cm}^3.
\]
The results here are applied to the mater with non-relativistic and
non-degenerate nucleons. The Fermi momentum of neutrons (protons)
$p_n\, (p_p)$ in a fully degenerate gas is written as \cite{bk01}

 \begin{equation}
\frac{p_{n,p}}{m_{n,p} c}\approx \left(\frac{\rho_{n,p}}{6.2\cdot
10^{15}{\rm g/cm}^3}\right)^{1/3},
 \label{nd}
\end{equation}
and $m_n\, c^2\,=931.5$\,MeV=$k\,5.4\cdot 10^{12}\,K$.    It is
evident from comparison with (\ref{nsph}), that nucleons are always
nonrelativistic at the neutrinosphere. For nonrelativistic,
nondegenerate nucleons their Fermi energy $E_{Fe}$ should be less
that $1.5 kT$. We have

 \begin{equation}
E_{Fe}=\frac{m_n c^2}{2} \left(\frac{\rho_{n}}{6.2\cdot 10^{15}{\rm
g/cm}^3}\right)^{2/3},\,\, E_{Fe}=\frac{3}{2}kT
 \label{nt}
\end{equation}
\[{\rm at} \,\,\, T=T_d=3.6\cdot 10^{10}\,
K\,\left(\frac{\rho_{n}}{6.2\cdot 10^{12}{\rm g/cm}^3}\right)^{2/3}.
\]
It follows from the comparison with (\ref{nsph}), that the
temperature at the neutrinosphere $T_{\nu}$ is always much larger
than $T_d$, so the approximation of nonrelativistic and
nondegenerate nucleons is always valid near the neutrinosphere.

\bigskip

{\bf Acknowledgments}

\medskip

The present work was partially supported by RFBR grants 11-02-00602,
RAN Program 'Origin, formation and evolution of objects of
Universe', and Russian Federation President Grant for Support of
Leading Scientific Schools NSh-3458.2010.2.

\appendix

\section{Calculations of Fermi functions by a generalized Gauss method}

Introducing a function

\begin{equation}
f(x,\alpha)=\frac{1}{e^{-x}+e^{-\alpha}},
\label{fun}
\end{equation}

\noindent let us represent the Fermi function $F_n(\alpha)$ as

\begin{equation}
F_n(\alpha)=\int_0^\infty f(x,\alpha)x^n e^{-x}dx.
\label{fi1}
\end{equation}
Gauss method of calculation of definite integrals suggest reducing it to an algebraic
relation, in which it is necessary to calculate the function $f(x)$ inside the integral in several number  $m$ of nodes, and calculate the sum of these values in fixed nodes $x_i$, with fixed coefficients $A_i$, $i\leq m$. For a polynomial function $f(x)$ this method gives an exact value of the integral when the power of the polynomial $p\leq 2m-1$. So there is a presentation

\begin{equation}
\int_{-1}^{1}f(x)dx=\sum_{i=1}^m A_if(x_i),
\label{fi2}
\end{equation}
where the value of $x_i$ and $A_i$ are calculated for number of nodes between  $m=2$ and $m=96$ in \cite{as}. To calculate improper integrals with infinite upper limits it is better to use a modified Gauss method, when the integrals with different asymptotic behavior are calculated as

 \begin{equation}
\int_0^\infty f(x)x^n e^{-x}dx=\sum_{i=1}^m A_{ni}f(x_{ni}).
\label{fi3}
\end{equation}
The values of $A_{ni}$ and $x_{ni}$ have been calculated in \cite{bkk67} for $m=5$, $n=0,\,1,\,2,\,3,\,4$, (see also \cite{bk01}). The values for $n=,2\,3,\,4$  are given in Tab.\ref{tab4}.
The Fermi functions from (\ref{int2}), with $f(x,\alpha)$ from (\ref{fun}), are represented as

\[F_2(\alpha)=0.52092\cdot f(1.0311,\alpha)+1.0667\cdot f(2.8372,\alpha)+\]
\begin{equation}
+0.38355\cdot f(5.6203,\alpha)+
\label{fi4}
\end{equation}
\[
+0.028564\cdot f(9.6829,\alpha)+  2.6271\cdot 10^{-4}\cdot f(15.828,\alpha),\]

\[F_3(\alpha)=1.2510\cdot f(1.4906,\alpha)+3.2386\cdot f(3.5813,\alpha)+\]
\begin{equation}
+1.3902\cdot f(6.6270,\alpha)+
\label{fi5}
\end{equation}
\[
+0.11904\cdot f(10.944,\alpha)+  1.2328\cdot 10^{-3}\cdot f(17.357,\alpha),\]

\[F_4(\alpha)=4.1856\cdot f(1.9859,\alpha)+12.877\cdot f(4.3417,\alpha)+\]
\begin{equation}
+6.3260\cdot f(7.6320,\alpha)+
\label{fi6}
\end{equation}
\[
+0.60475\cdot f(12.188,\alpha)+  6.8976\cdot 10^{-3}\cdot f(18.852,\alpha),\]
After calculation of $I_2$ and $J_2$ in (\ref{int2}), the values of $Z$, $\mu$, $T$, $\rho$, are found as

\begin{equation}
Z=3\pi^2 \frac{\rho}{m_p}\left(\frac{\hbar c}{kT}\right)^3=(\beta^3+\pi^2\beta)\left(1+\frac{I_2}{J_2}\right),
\label{fi7}
\end{equation}
\[\mu=1+\frac{I_2}{J_2},\quad T=\frac{\Delta}{k x_0}, \quad \rho=\frac{Z m_p}{3 \pi^2} \left(\frac{kT}{\hbar c}\right)^3.\]

\begin{table}
\begin{center}
\caption{Roots and coefficients for calculating integrals (\ref{fi3})
with $m=5,\,\, n=2,3,4$, from \cite{bk01}.}
\vspace{3mm}
\label{tab4}
 \footnotesize{
 \begin{tabular}{clllll}
\hline
  Roots $x_i$ &&&&& \\
  and          & $n=2$ & $n=3$ & $n=4$  \\
  coefficients $A_i$ &&& \\
\hline
 &&&&& \\
 $x_1$  & 1.0311     & 1.4906     & 1.9859     \\
 $x_2$  & 2.8372     & 3.5813     & 4.3417     \\
 $x_3$  & 5.6203     & 6.6270     & 7.6320     \\
 $x_4$  & 9.6829     & 10.944     & 12.188    \\
 $x_5$  & 15.828     & 17.357     & 18.852     \\
 $A_1$  & 0.52092    & 1.2510     & 4.1856     \\
 $A_2$  & 1.0667     & 3.2386     & 12.877     \\
 $A_3$  & 0.38355    & 1.3902     & 6.3260     \\
 $A_4$  & 0.028564   & 0.11904    & 0.60475    \\
 $A_5$  & 2.6271(-4) & 1.2328(-3) & 6.8976(-3) \\
 &&&&& \\
\hline
\end{tabular}}
\end{center}
\end{table}


\begin{thebibliography}{99}

\bibitem{as}
 Abramowitz, M. and  Stegun, I.A. (eds) (1964).
Handbook of mathematical functions
with formulas, graphs and mathematical tables.
National bureau of standards
applied mathematics series - 55

\bibitem{bk01} Bisnovatyi-Kogan, G.S. (2001).
Stellar Physics.
Vol.1. Fundamental concepts and stellar equilibrium, Springer.

\bibitem{bkk67} Bisnovatyi-Kogan, G.S., Kazhdan, Ya.M. (1966).
 Astron. Zh.  {\bf 43},   761 (Soviet Astronomy {\bf 10}, 603, 1967)

\bibitem{chech69}
Chechetkin, V.M. (1969).  Astron. Zh.  {\bf 46},   202 (Soviet Astronomy {\bf 13}, 153, 1969)

\bibitem{in65} Imshennik, V. S., Nadyozhin, D. K. (1965). Astron. Zh.  {\bf 42},   1154 (Soviet Astronomy {\bf 9}, 896, 1966)

\bibitem{inp67} Imshennik, V. S., Nadyozhin, D. K., Pinaev, V. S. (1967). Astron. Zh.  {\bf 44}, 768 (Soviet Astronomy {\bf 11}, 617, 1968)

\bibitem{nad74} Nadyozhin, D. K. (1974). Nauchnyie Informatzii
 Astronomicheskogo Soveta AN SSSR (Scientific Information of the Astonomical
 Council of the Academy of Sciences of the USSR), Issue 32, 3

\bibitem{n78} Nadyozhin, D. K. (1978). Ap\&SS {\bf 53}, 131

\bibitem{rho50}
Rhodes, P. (1950).  Proceedings of the Royal Society of London. Series A, Mathematical and Physical Sciences {\bf 204:1078}. 396

\bibitem{prd05}
Ye-Fei Yuan (2005). Physical Review {\bf D 72}, 013007

\end{thebibliography}
\end{document}